\documentclass[%
 reprint,
 superscriptaddress,
showpacs,
 amsmath,amssymb,
 aps,
]{revtex4-1}

\usepackage{graphicx,epsf}
\usepackage{dcolumn}
\usepackage{bm}
\usepackage{braket}
\usepackage{xcolor}

\newcommand{\red}{ }                

\usepackage[normalem]{ulem}

\begin{document}

\title{Conductance features of core-shell nanowires determined by the
internal geometry}

\author{Miguel Urbaneja Torres}
 \affiliation{School of Science and Engineering, Reykjavik University, 
              Menntavegur 1, IS-101 Reykjavik, Iceland}   

\author{Anna Sitek}
\affiliation{School of Science and Engineering, Reykjavik University, 
              Menntavegur 1, IS-101 Reykjavik, Iceland}
\affiliation{Department of Theoretical Physics,
              Faculty of Fundamental Problems of Technology,
              Wroclaw University of Science and Technology,
              Wybrze{\.z}e Wyspia{\'n}skiego 27, 50-370 Wroclaw, Poland}

\author{Sigurdur I. Erlingsson}
 \affiliation{School of Science and Engineering, Reykjavik University, 
              Menntavegur 1, IS-101 Reykjavik, Iceland}
 
\author{Gunnar Thorgilsson}
 \affiliation{School of Science and Engineering, Reykjavik University, 
              Menntavegur 1, IS-101 Reykjavik, Iceland}

\author{Vidar Gudmundsson}
 \affiliation{Science Institute, University of Iceland, Dunhaga 3, 
              IS-107 Reykjavik, Iceland}

\author{Andrei Manolescu}
 \affiliation{School of Science and Engineering, Reykjavik University, 
              Menntavegur 1, IS-101 Reykjavik, Iceland}

\begin{abstract}
We consider electrons in tubular nanowires with prismatic geometry
and infinite length. Such a model corresponds to a core-shell nanowire with 
an insulating core and a conductive shell.  In a prismatic shell the lowest 
energy states are localized along the edges (corners) of the
prism and are separated by a considerable energy gap from the states
localized on the prism facets.  The corner localization is robust
in the presence of a magnetic field longitudinal to the wire.  If the
magnetic field is transversal to the wire the lowest states can be
shifted to the lateral regions of the shell, relatively to the direction
of the field. These localization effects should be observable in transport
experiments on semiconductor core-shell nanowires, typically with
hexagonal geometry.  We show that the conductance of the prismatic
structures considerably differs from the one of circular nanowires. The
effects are observed for sufficiently thin hexagonal wires and become
much more pronounced for square and triangular shells. To the best of
our knowledge the internal geometry of such nanowires is not revealed
in experimental studies. We show that with properly designed
nanowires these localization effects may become an important resource
of interesting phenomenology.
\end{abstract}


\maketitle

\section{\label{sec:introduction} Introduction}

The fabrication of various types of nanostructures allows for the design of
systems with controllable properties of the electronic states.  Amongst such
systems are core-shell nanowires, which are radial heterojunctions of
two, or even more, different semiconductor materials.  Typically a
central material (core) is surrounded by an outer layer (shell).
The length of such nanowires is usually of the order of microns, whereas
the diameter is between a few tens to a few hundreds of nanometers.
An interesting aspect of this structure is the transversal geometry.
Semiconductor core-shell nanowires, most often based on III-V materials,
are almost always prismatic, and rarely cylindrical.  The typical shape
of the cross section is hexagonal 
\cite{Rieger12,Blomers13,Haas13,Plochocka13,Jadczak14,Pemasiri15}.
Interestingly, other prismatic geometries can also be
achieved, like square \cite{Fan06} or even triangular
\cite{Qian04,Qian05,Baird09,Heurlin15,Dong09,Yuan15}.
The present art of manufacturing allows even for etching out the core  
and obtaining prismatic semiconductor nanotubes with vacuum inside
\cite{Haas13,Rieger12}.

In general, the specific geometry of the core-shell nanowires has not 
been studied much experimentally. It is either seen as
a natural outcome, when the nanowires are hexagonal, or as a curiosity,
when they are of different shapes. The main interest is rather related
to the materials used, the quality of the nanowires (like defect free),
a specific size of the core or the shell, specific length, etc.  
Still, in particular, the triangular wires showed very interesting features such as a 
broad range of emitted wavelengths at room temperature \cite{Qian08,Qian05} or
diode characteristics\cite{Dong09},
but such properties were not really associated with the shape 
of the cross section and/or with particle localization.
\red{Remarkably, triangular nanowires possess intrinsic
polarization effects that may break the three-fold geometric symmetry of the
cross-section \cite{Wong11}.}

Theoretically, the prismatic geometry of the nanowire, and especially of the 
shell itself, is a unique, and a very important feature, which can lead to very 
interesting and rich physics.  If the materials, the geometry, the doping, 
and the shell thickness are properly adjusted,   
the shell becomes a tubular conductor with edges, and each 
edge may behave like a quasi-one-dimensional channel. 
\red{A pioneer theoretical work has shown that carriers in prismatic shells
can form a set of quasi-one-dimensional quantum channels localized at the prism edges
\cite{Ferrari09b}. More recently it has been} 
predicted that such a prismatic shell can host several Majorana states
which may interact with each other \cite{Manolescu17,Stanescu18}.
Another recent prediction is that a magnetic field transversal to a tubular
nanowire, either made of normal semiconductors, or of a topological insulator material,  
may induce the sign reversal of the electric current generated by a 
temperature gradient \cite{Erlingsson17,Thorgilsson17,Erlingsson18}. 

In this paper we consider prismatic tubes as models of core-shell
nanowires with an insulating core and a conductive shell.  Our models are
also applicable to uniform nanowires (i.e.\ not core-shell) which have
conductive surface states due to the Fermi level pinning \cite{Heedt16}.
However, we are not addressing the nanowires built from topological
materials, but we use in our calculations a Schr\"odinger Hamiltonian.
In our models the cross section of the prismatic shell is a narrow 
polygonal ring (hexagonal, square, or triangular), i.e.,\ with lateral
thickness much smaller than the overall diameter of the nanowire.

In this geometry the electrons with the lowest energies are localized in 
the corners of the polygon and the electrons in the next layer of energy states
are localized on the sides \cite{Bertoni11a,Royo13,Royo14,Royo15,Fickenscher13,Shi15}.
The corner and side states are energetically separated by an interval 
that depends on the geometry and on the aspect ratio of the polygon, it
increases with decreasing the shell thickness or the number of corners,  
and it can become comparable or larger than the energy corresponding 
to the room temperature \cite{Sitek15,Sitek16}.  Hence, such structures
can contain a well-separated subspace of corner states, with sharp
localization peaks, and potentially robust to many types of perturbations.
On the contrary, if the shell is relatively thick with respect to
the diameter of the wire, the corner localization broadens and the
polygonal structure has less effect on the electron distribution
\cite{Sitek16ICTON}.

To the best of our knowledge, experimental results with features related to the 
internal geometry of the nanowire do not exist, or are very rare. 
One investigation, using inelastic light scattering, indicated the coexistence
of one- and two-dimensional electron channels, along the edges and facets, respectively, 
of GaAs core-shell nanowires \cite{Funk13}.
In transport experiments one can mention here the detection of flux periodic oscillations of the conductance
in the presence of a magnetic field longitudinal to the nanowire \cite{Gul14}, 
which indicates the radial localization of the electrons in the shell.
Or, flux periodic oscillations with the magnetic field perpendicular to the 
nanowire, due to the formation of snaking states on the sides of the tubular
conductor \cite{Heedt16,Manolescu16}. Nevertheless, \red{experimental} 
results indicating the presence
of corner or side localized states, or the energy gap between them, or other
details implied by the prismatic geometry of the shell, are not reported.

The intention of this paper is to predict specific features of 
the conductance of core-shell nanowires when the electronic transport occurs 
within the shell, determined by the prismatic geometry of the nanowire.
Such features, which could be experimentally tested, can reveal to what
extent the electronic states are influenced by the polygonal geometry,
and if not, to what extent the quality of the nanowire needs to be improved
in order to achieve a robust corner localization. 

Next, in Section II we describe our model and methodology,  in Section III
we discuss the transverse modes and the expected conductance steps, in 
Section IV we consider a magnetic field longitudinal to the nanowire, 
in Section V a perpendicular magnetic field and finally in 
Section VI we summarize the conclusions.

\section{\label{sec_model} Model and methods}

We analyze a system of non-interacting electrons, confined in a prismatic
shell with polygonal cross section, in the presence of a uniform magnetic field
${\bf B}=(B_x,B_y,B_z)$.  The axes $x$ and $y$ are chosen perpendicular to the 
shell and the axis $z$ is longitudinal. The Hamiltonian can be decomposed as 
\begin{eqnarray}
\label{hamiltonian}
 H = H_t + H_l + H_s \ ,
\label{Htot}
\end{eqnarray}
where the terms $H_t$, $H_l$, and $H_s$ correspond to the transverse, the 
longitudinal, and the spin degrees of freedom, respectively.

The transverse Hamiltonian is, 
\begin{equation}
\label{hamiltonian_t}
 H_t = \frac{(-i\hbar\partial_x+eA_x)^2+(-i\hbar\partial_y+eA_y)^2}
{2m_{\rm eff}}
-e\bm{E}\cdot\!\bm{r}\ ,
\end{equation}
where $m_{\rm eff}$ is the effective electron mass in the shell material, 
${\bf A}=(A_x,A_y,A_z)=(-yB_z/2,\ xB_z/2,\ yB_x-xB_y)$ is the vector potential,
and ${\bf E}=(E_x,E_y,0)$ is an 
external electric field perpendicular to the wire.  The transverse Hamiltonian
depends only on the longitudinal magnetic field $B_z$. 

\begin{figure}[t]
\centering
\includegraphics[scale=0.62]{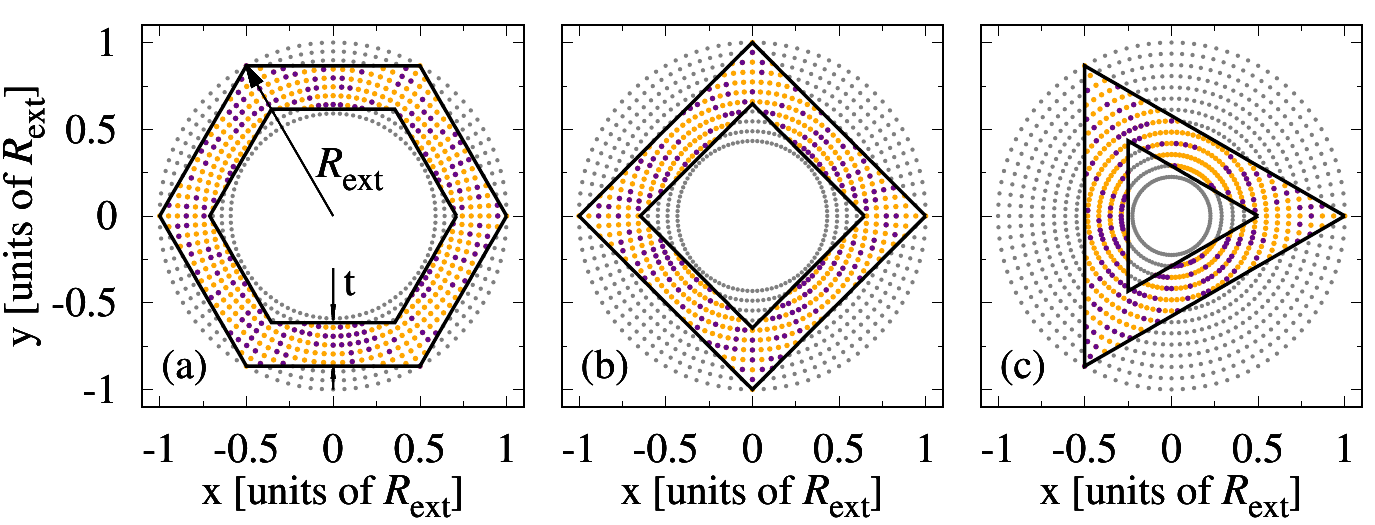}
 \caption{The cross sections of the prismatic 
          shells are defined by applying boundaries 
          on a circular ring discretized in polar 
          coordinates. Only the lattice points inside the polygonal shell 
          (shown in yellow) are used in the transverse 
          Hamiltonian (1), where we can include a chosen percentage of random-strength
impurities (purple points). 
          $R_{\mathrm{ext}}$ and $t$ indicate the nanowire radius and 
          shell thickness, respectively.
          For clarity the figures show only a subset of lattice points,
          whereas the numerical calculations are performed with 
          6000-10000 points, depending on the polygon: 
          (a) hexagon, (b) square, (c) triangle.}
\label{Fig_Sample}
\end{figure}

Technically, the Hamiltonian (\ref{hamiltonian_t})
is restricted to a lattice of points that covers the cross section of the shell.
In order to define the lattice
we begin with a circular disk which is discretized in polar coordinates
\cite{Daday11}.  Next, we enclose the polygonal shell within this area
and exclude all lattice points situated outside the shell, as shown in 
Fig.\ \ref{Fig_Sample}.  This method allows us to 
describe both symmetric and non-symmetric polygonal shells without the need
of adapting the background grid geometry to the specific polygon
or redefining Hamiltonian matrix elements \cite{Sitek15,Sitek16}.  
As a second method we also used the Kwant software \cite{Groth14} 
and obtained the same numerical results for the matrix elements, this time
using a triangular grid.

The longitudinal Hamiltonian is 
\begin{equation}
\label{hamiltonian_l}
 H_l = \frac{(-i\hbar\partial_z+eA_z)^2} {2m_{\rm eff}}\ ,
\end{equation}
which depends on the magnetic
field transverse to the nanowire ${\bf B}_{\perp}=(B_x,B_y)$, which
can be chosen at different angles relatively to the corners or sides of
the shell.  

And finally the spin Hamiltonian is 
\begin{equation}
 H_s = -g_{\rm eff}\ \mu_{\rm B}\ {\bm \sigma}\cdot{\bm B}\ ,
\end{equation}
where $g_{\rm eff}$ is the effective g factor, $\mu_{\rm B}$ is Bohr's
magneton,  ${\bm
\sigma}=(\sigma_x,\sigma_y,\sigma_z)$ denote the Pauli matrices,
and ${\bm B}$ is the total magnetic field.

In our numerical calculations we first calculate the eigenstates of the 
transverse Hamiltonian, $H_t \ket{a}=\epsilon_a \ket{a}$ ($a=1,2,3,...$),
and obtain the corresponding eigenvectors in the position representation,
$\ket{a} = \sum_{q}\psi(q,a)\ket{q}$, where $|\psi(q,a)|^{2}$, is the
localization probability on the lattice site $q=(x_q,y_q)$.  Next, we
retain only a set of low-energy transverse modes, and together with the
plane waves in the $z$ direction $\ket{k} = \exp(ikz)/\sqrt{L}$, $k$
being the wave vector and $L$ the (infinite) length of the nanowire,
and with the spin states $s=\pm 1$, we form a basis in the Hilbert space
of the total Hamiltonian, $\ket{aks}$. In the absence of the magnetic
field transverse to the wire (${\bf B}_{\perp}=0$), the kets $\ket{aks}$
are eigenvectors of the total Hamiltonian, with eigenvalues 
$E_{aks} = \epsilon_{a} + \hbar^2k^2/2m_{\rm eff}-g_{\rm eff}\mu_{\rm B} sB_z$.
If ${\bf B}_{\perp}\neq 0$ the transverse motion becomes dependent
on $k$, and then we diagonalize the total Hamiltonian for a discretized series of $k$
values, to obtain its eigenvalues $E_{mks}$ ($m=1,2,3,...$), and its
eigenvectors $\ket{mks}$ expanded in the basis $\ket{aks}$.

The next step of our calculations is to evaluate the electric current
along the nanowire in the presence of a voltage bias.  The operator
describing the contribution of an electron which is localized at point
$\bm{r}_{0}$, to the total charge-current density observed in a spatial
point $\bm{r}$, is defined as
\begin{equation}
\label{current}
\bm{j}(\bm{r},\bm{r}_{0}) = \frac{e}{2}[\delta(\bm{r}-\bm{r}_0)\bm{v} + \bm{v}\delta(\bm{r}-\bm{r}_0)] \ ,
\nonumber 
\end{equation}
where \red{$\delta(\bm{r}-\bm{r}_0)$ is the particle-density operator and
$\bm{v}(\bm{r}_0) = \frac{i}{\hbar}[H,\bm{r}_0]$ the velocity
operator \cite{Messiah}.}  In our case we need only the component along the nanowire, 
$v_z=(p_z+eA_z)/m_{\rm eff}$. 
We can define the expected value of the total charge current flowing in 
the positive or negative direction along the nanowire (i.e. in the $z$ direction) 
as
\begin{equation}
\label{expected_current}
I_{\pm}\!=\!\int\!\!\left[\sum_{mks} 
\! {\cal F} \! \left(\frac{E_{mks}-\mu_{\pm}}{k_{B}T}\right) 
\! \braket{mks|\bm{j}(\bm{r},\bm{r}_{0})|mks}\!\right]\!\!\mathrm{d}{\bm r} .\!
\end{equation}
The integration is performed over the cross section of the shell,
practically as a summation over all lattice sites ${\bm r}_{q=}(x_q,y_q)$,
whereas the scalar product included in the square brackets is an
integration over the electrons position $\bm{r}_0$.  ${\cal F}(w) =
1/[\mathrm{exp}(w)+1]$ represents the Fermi function, 
$T$ is the temperature, and $k_B$ Boltzmann's
constant.  Here $\mu_{\pm}$ are the chemical potentials associated
with electrons having positive or negative velocity along the nanowire.
Obviously, in equilibrium $\mu_{+}=\mu_{-}$ the corresponding currents
compensate each other and the total current $I=I_{+}-I_{-}$ is zero.

To generate a current along the nanowire we create an
imbalance between the states with positive velocity, i.e., $\partial
E_{mks}/ \partial k > 0$, and negative velocity, i.e.,  $\partial E_{mks}/
\partial k<0$, by considering in Eq.\ (\ref{expected_current}) different
chemical potentials, $\mu_{+}$ and $\mu_{-}$, respectively. The current
is thus driven along the nanowire by the potential bias $eV = \mu_{+}
- \mu_{-}$.  This procedure, well established in ballistic transport
theory \cite{Datta}, allows us to calculate the $I-V$ characteristic
and the conductance $G = I/V$ in the small bias limit.

To include the effect of disorder on the conductance we consider a nanowire 
with a scattering region of a finite length containing a random distribution of 
impurities, and we assume elastic electron-impurity collisions.
In this case the impurities can be represented by an
extra random term in the total Hamiltonian (\ref {Htot}).  The 
current and the conductance can be calculated using the well known
concept of transmission function. Here we compute the transmission function by 
using the Fisher-Lee formula \cite{Fisher81} and the recursive Green's 
function method \cite{Ferry97}. A summary of the method for cylindrical 
geometry can be found in Ref.\ \cite{Erlingsson17} (Supplemental Material).

We also address in our paper the case of a long nanowire, much
longer than the scattering length, when the transport is far from
ballistic.  In principle this case can be treated with the recursive
Green's functions method, by including a large number of impurities,
but it becomes computationally very expensive.  As an alternative approach 
we will consider in this case a nanowire of infinite length, and we will use
the traditional Kubo formula, which allows one to calculate the
conductivity by performing a statistical average over impurity 
configurations \cite{Doniach98}: 
\begin{equation}
\begin{split}
& \sigma_{zz}  = \   \frac{he^2}{V} \!
\int \! dE \left[-\frac{\partial \cal F} {\partial E}\right]\ \times \\
& \!\! \sum_{\substack{m_1 k_1 s_1 \\ m_2 k_2 s_2} } \! 
|\braket{m_1 k_1 s_1 |v_z|m_2 k_2 s_2}|^2  
 A_{m_1 k_1 s_1}(E) \  A_{m_2 k_2 s_2}(E) \ ,
\label{kuboc}
\end{split}
\end{equation}
where $V$ is the volume of the conductor and $A_{mks}(E)$ is the spectral 
function which represents the broadening of the energy level $E_{mks}$ due to 
the disorder.  In this study we model the spectral
function for each energy level $E_{nks}$ as a Gaussian function, 
\begin{equation}
\label{spectral_f}
A_{mks}(E)=\frac{1}{\sqrt{2\pi}\Gamma} e^{-\frac{(E-E_{mks})^2}{2\Gamma^2}}\ ,
\end{equation}
whose width parameter $\Gamma$ represents the disorder energy.  This model
of spectral function was used in the past for describing the impurity effects
in the two-dimensional electron gas \cite{Ando85}, but also more recently to incorporate
the effect of the electrodes on the density of states in molecular nanowires \cite{Tada04}. 

In the next sections we will show results for the conductance
of core-shell nanowires, related to their internal geometry, in different
situations, using the computational methods mentioned above.

\section{Transverse modes for zero magnetic field}

In all the following examples the external radius of the polygonal shell,
which is the distance from the center to one corner, as indicated in
Fig.\ \ref{Fig_Sample}(a), is  $R_{\mathrm{ext}} = 30$ nm, whereas the
side thickness $t$ varies between 2-8 nm.  The numerical calculations
were performed for InAs bulk parameters, which are $m_{\rm eff} =
0.023 m_{\mathrm{e}}$ ($m_{\mathrm{e}}$ being the electron mass of a
free electron), and $g_{\rm eff}=-14.9$.

\begin{figure}[t]
\centering
\includegraphics[scale=0.10]{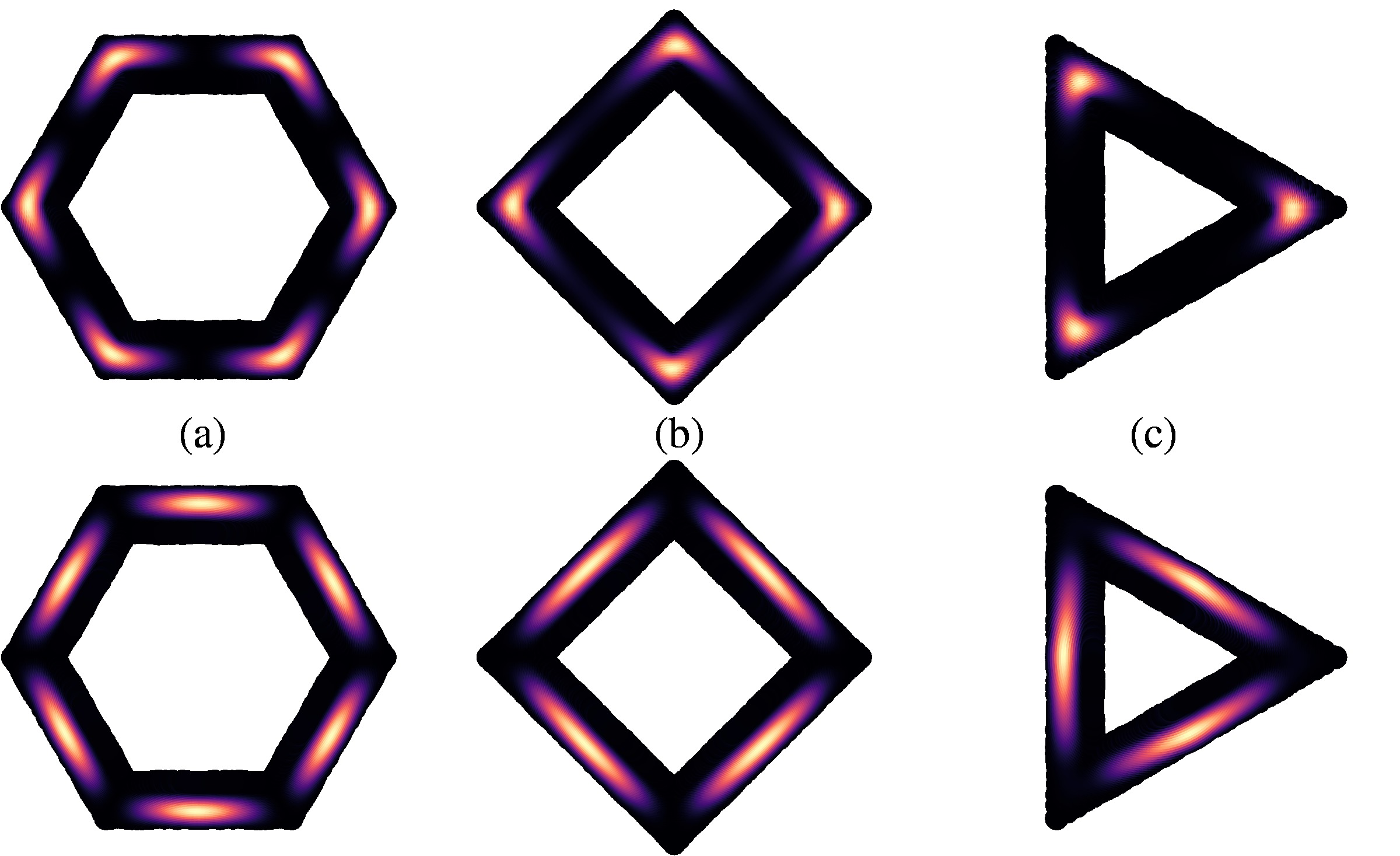}    
\caption{Single-particle probability distributions associated with the low-energy
transverse modes of (a) hexagonal, (b) square, and (c) triangular
shells. The top figures illustrate the corner states, which include the
ground state, and the bottom figures illustrate the next, higher-energy
group of modes, localized on sides. The energy separation between corner and side 
modes is indicated in Fig.\ \ref{fig_sides}.}
\label{fig_single_loc}
\end{figure}

Because of the polygonal cross section of the shell the lowest transverse
modes are localized in the corners of the polygons
while the probability distributions corresponding to higher modes have
maxima on the sides \cite{Sitek15}, as illustrated in Fig.\ \ref{fig_single_loc}. 
In three dimensions the corner states form 1D conductive channels
along the edges of the prismatic shell \cite{Shi15}.  For each polygon there are $2N$
corner states, where $N=6,4,3$ is the number of corners (or sides) and the
factor 2 accounts for the spin.  The energy dispersion of corner states 
decreases with the number of corners.  For example,
for $t=6$ nm shell thickness these states fit into the intervals of
15.3 meV  for the the hexagon, 3.1 meV for the square, and 0.2 meV for the triangle
(measured from the ground state).  For each geometry,
above the group of corner states, there is another group of $2N$ states,
localized on the the sides of the polygons (the lower part of  
Fig.\ \ref{fig_single_loc}). For the present radius (30 nm) and side thickness
(6 nm) the energy dispersion within each group of corner and side states 
is exceeded by the energy interval (or gap) between these groups, 
which are  $\Delta_{\rm h}=21.2$ meV for hexagon, $\Delta_{\rm s}=34.4$
meV for square and $\Delta_{\rm t}=87.1$ meV for triangle. 

These energy gaps make a big difference between a polygonal and a
cylindrical shell.  In the latter case, if $t\ll R_{\rm ext}$, one
expects the transverse energies $\varepsilon_m=\hbar^2m^2/2m_{\rm
eff}R_{\rm ext}^2$, with $m=0,\pm 1, \pm 2, ...$ the quantum number
of the angular momentum, i.e., with energy intervals uniformly increasing
as $m^2$.  In addition, in the cylindrical case (not shown in Fig.\
\ref{fig_single_loc}), at zero magnetic field, the transverse modes are
four fold degenerate, both spin and orbital, except the ground state
which is only two fold, spin degenerate. For symmetric polygons, because
of the reduced, discrete symmetry, the orbital degeneracy is lifted
between consecutive groups of $2N$ transverse states. In particular,
for symmetric shells, the \red{sequence of degeneracy orders (2-fold or 4-fold) 
of the corner/side states, with increasing the energy, is 2442/2442 for the hexagon, 242/242 for the square,
and 24/42 for the triangle.} 

\begin{figure}[t]
 \centering
\includegraphics[scale=0.62]{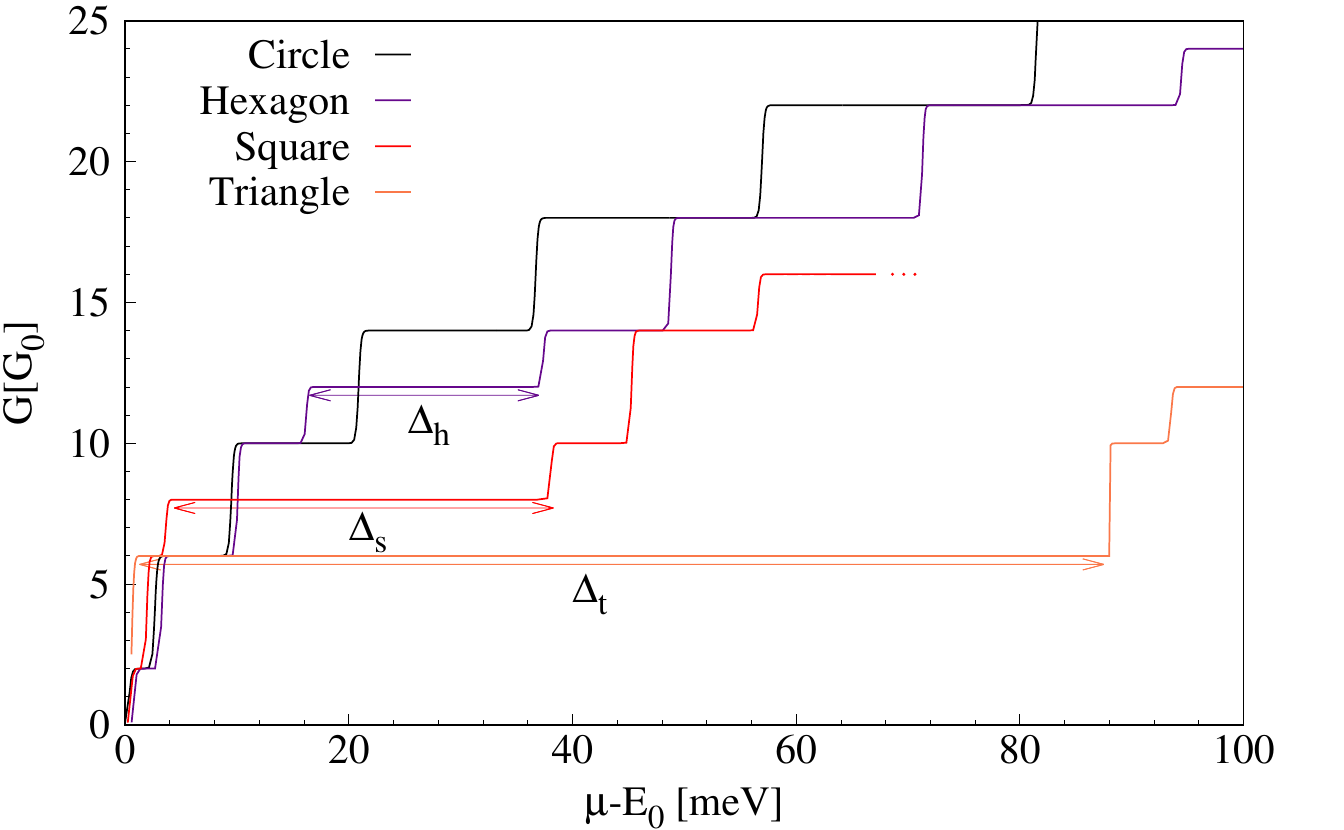}
\caption{Conductance steps expected for  
tubular nanowires of circular and polygonal cross sections, 
versus the chemical potential $\mu$ adjusted by subtracting the 
the ground state energy $E_0$.
$\Delta_{\rm h,t,s}$ indicate the 
energy separation between the corner and side modes for hexagon, square, and triangle, 
respectively.  The conductance is represented in units of $G_0=e^2/h$. Here $t = 0.2R_{\rm ext}=6$ nm}
\label{fig_sides}
\end{figure}

In Fig.\ \ref{fig_sides} we compare the conductance steps expected in
the cylindrical and in the three prismatic geometries, when the chemical
potential $\mu$ is varied.  Here we assume ballistic transport
and symmetric shells.  The conductance is derived with Eq.\ (\ref{expected_current}), 
using a small bias and temperature $T=1$ K. 
The steps obtained for circular and hexagonal
structures are visibly different, with a extra plateau at $12G_0$ for
the hexagon, ($G_0=e^2/h$) corresponding to the gap between corner and
side states.  These plateaus increase dramatically for the triangular and
square geometries, at $8G_0$ and at $6G_0$, respectively.  
For the hexagonal shell the corner states are visible
as steps at $2G_0$, $6G_0$, and $10G_0$, similar to the circular case.
For the square, instead, we see only some shoulders indicating the
corner states at $2G_0$ and $6G_0$, and for the triangular case they are not resolved.

\begin{figure}[t]
 \centering
 \includegraphics[scale=0.62]{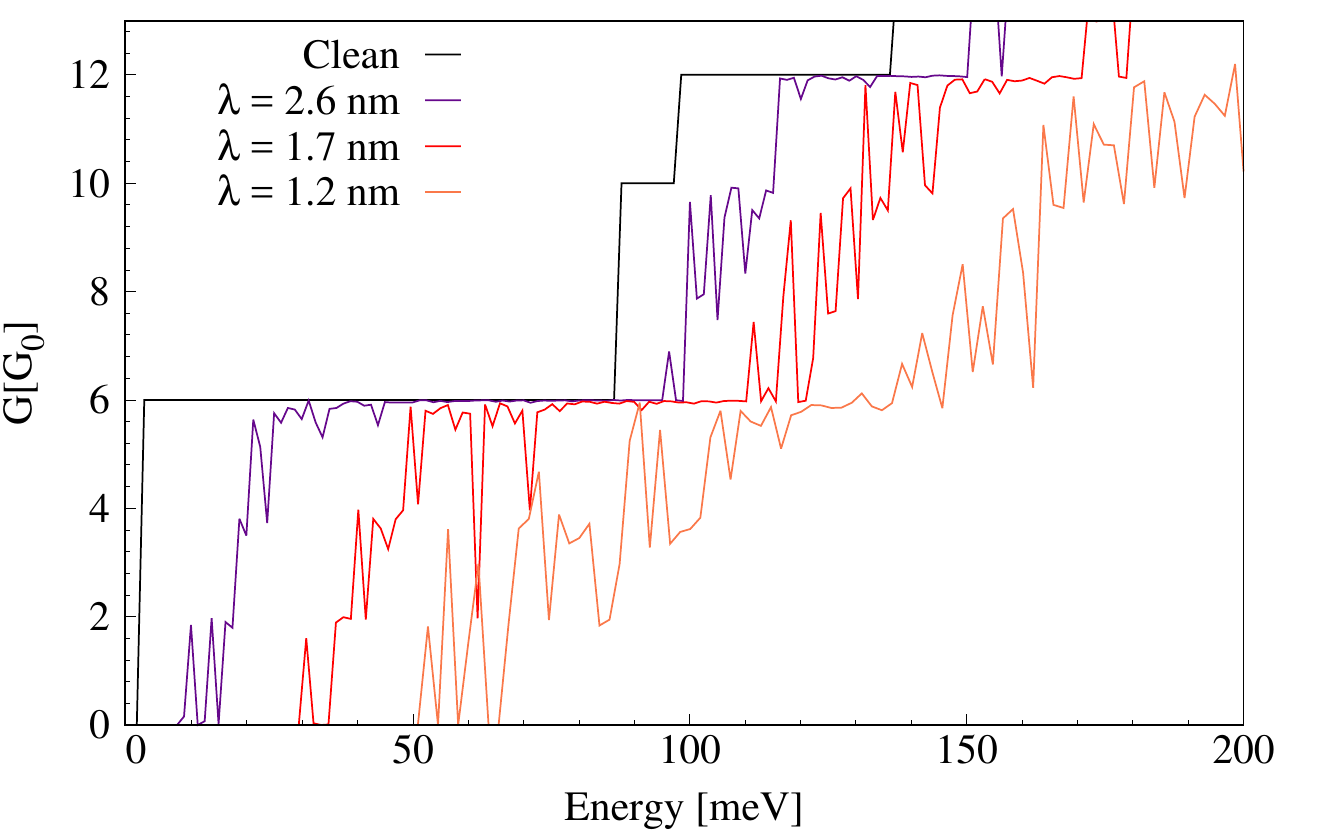}
\caption{Conductance steps expected for the triangular nanowire for different impurity concentrations  
characterized by the mean distance $\lambda$ between the nearest neighbors.
The associated potentials are repulsive and of a random strength between 0 and 0.5 meV.
}
\label{fig_impurities}
\end{figure}

In order to obtain the conductance of a nanowire 
with impurities we consider scattering centers,  with random characteristic energy
between 0 and 0.5 meV, at random locations 
within the tubular nanowire. We calculate the conductance, this time with
the recursive Green's function approach, for the example of the triangular
geometry, at zero temperature. The results are shown in Fig.\ \ref{fig_impurities} for
a nanowire of 54.2 nm length with different impurity concentrations
corresponding to a mean distance between the nearest neighbors of 2.7 nm, 1.6 nm and 
1.2 nm.  The impurity configurations are fixed in these cases, such that this
situation corresponds to a specific mesoscopic sample with individual
randomness.  The impurity concentration is increased until the largest plateau,
corresponding to the gap between the corner and side states, becomes nearly indistinguishable.
This impurity concentration can offer a hint on how dirty a nanowire can be
such that the conductance steps cannot be detected in the experiments.  

\begin{figure}[t]
 \centering
 \includegraphics[scale=0.62]{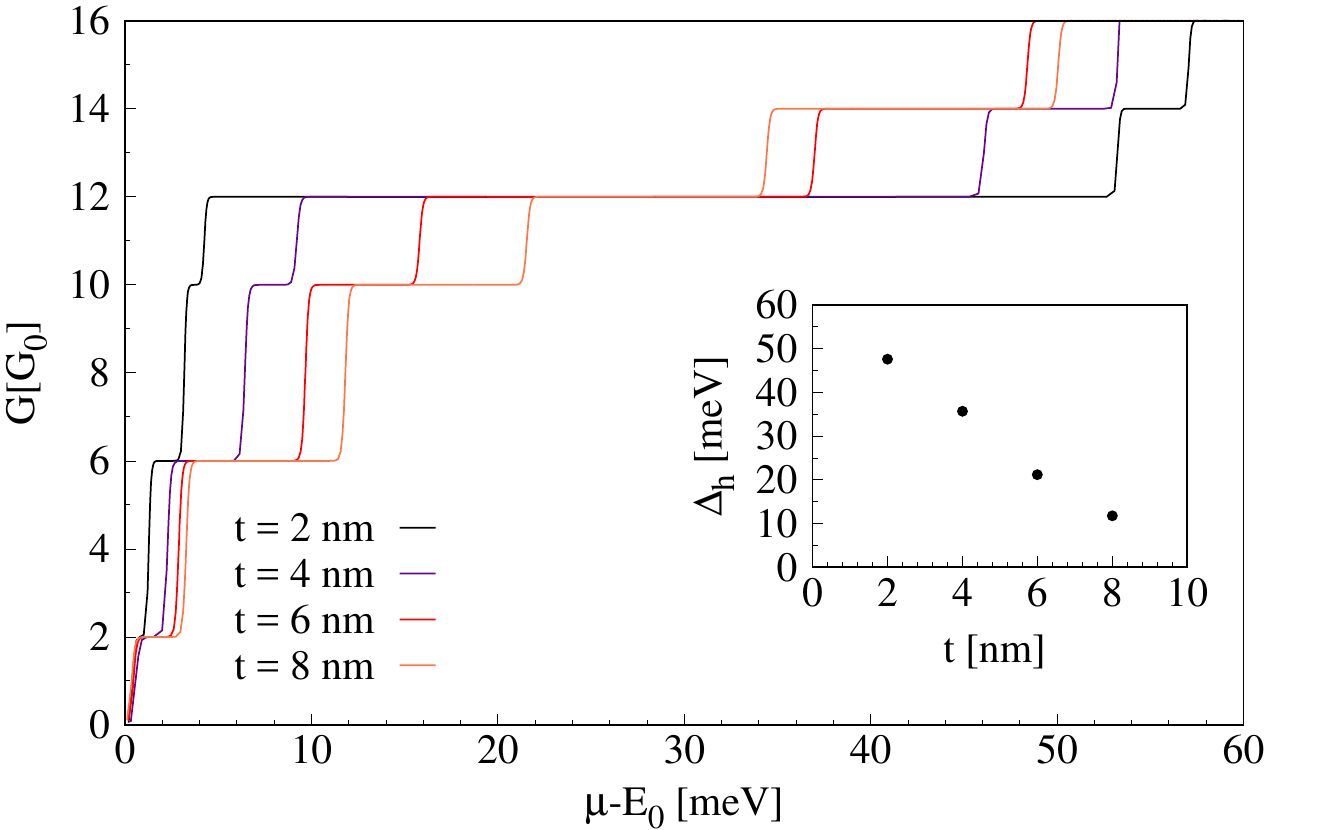}
 \caption{Dependence of the conductance plateaus on the side thickness
 $t$ for the hexagonal shells. 
 Decreasing the side thickness leads to a significant increase of 
 the plateau between the corner and side states, at $12G_0$, 
 and to a decrease of all the other plateaus.}
\label{fig_sides_comp}
\end{figure}

We now return to the clean nanowire to discuss the effect of the side thickness
on the hexagonal geometry. With the parameters used in Fig.\ \ref{fig_sides} the 
plateaus corresponding to side modes, i.e. above $12G_0$, are comparable or larger
than $\Delta_h$.  However, the relative magnitude of the energy intervals,
i.e., of $\Delta_h$ relatively to the dispersion of the corner and side
states, increases with reducing the aspect ratio of the polygon, i.e.,
$t/R_{\rm ext}$ \cite{Sitek16}.  As a result, with reducing the thickness
parameter $t$, while $R_{\mathrm{ext}}$ is kept constant, the plateau
at the transition between corner and side states becomes
much more prominent. At the same time, the ratio between $\Delta_h$ and
the other energy intervals rapidly increases such that the plateaus in
the corner domain become negligible relatively to the main one of width
$\Delta_h$, as shown in Fig.\ \ref{fig_sides_comp}.

\begin{figure}[t]
 \centering
 \includegraphics[scale=0.5]{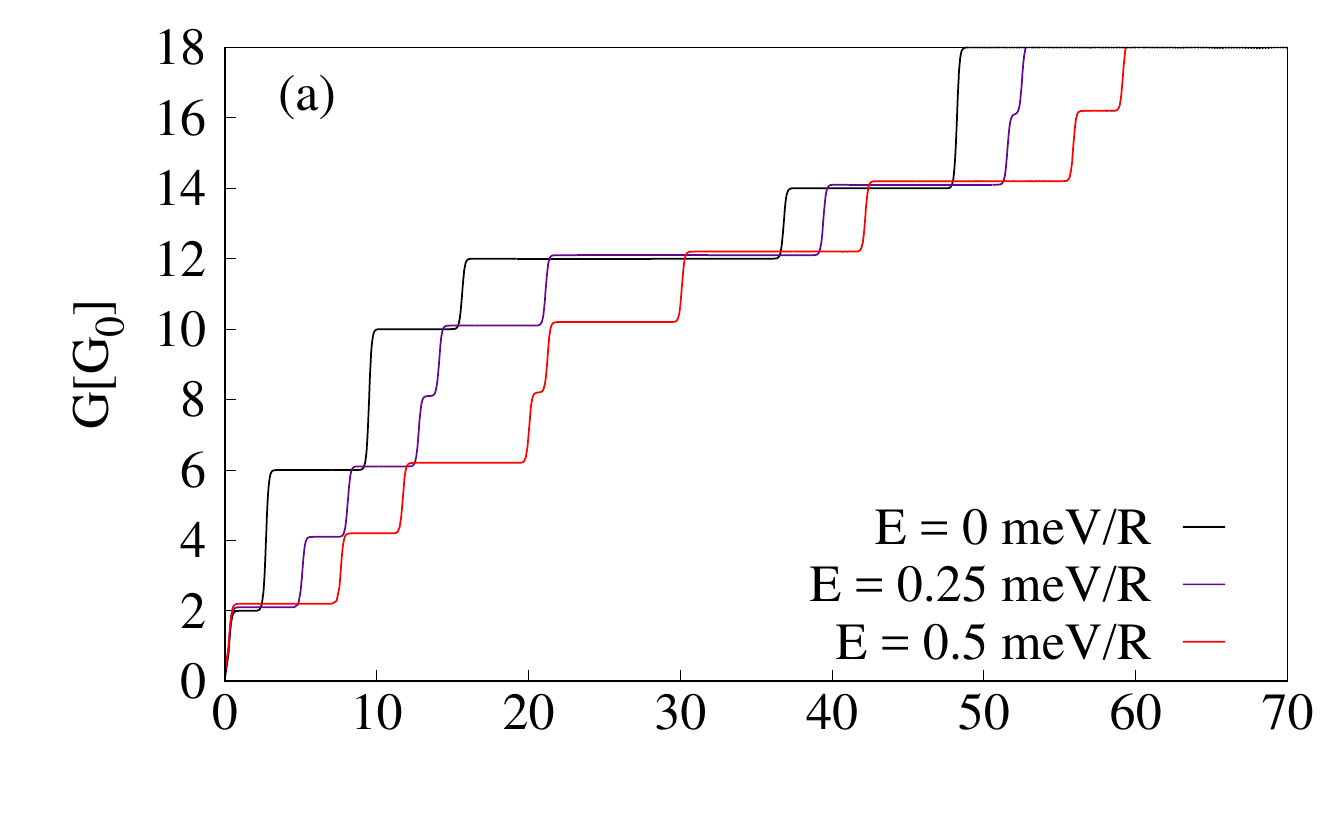}
 \includegraphics[scale=0.5]{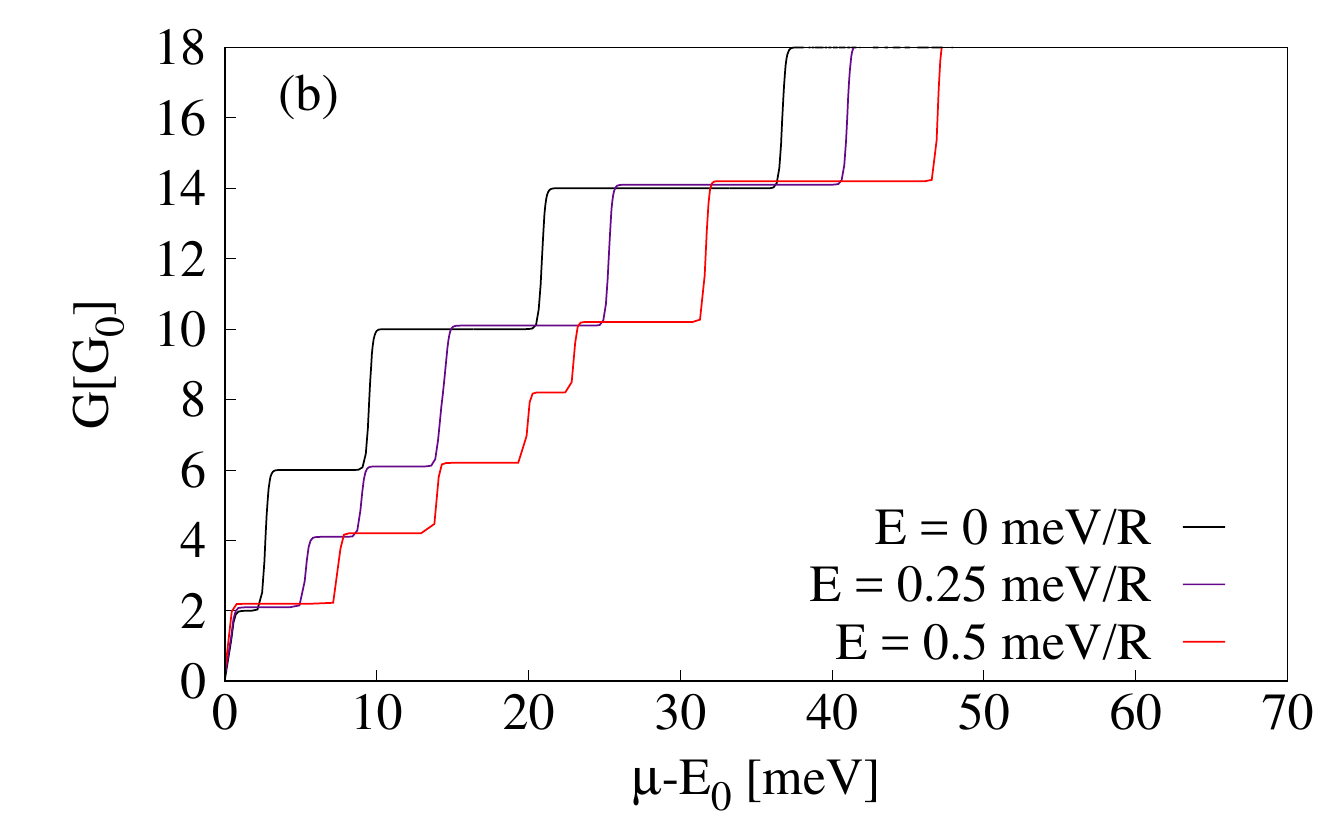}
\caption{Conductance steps in the presence of an asymmetry induced
by an electric field perpendicular to the nanowire, along the $x$-axis, for (a) the hexagonal case and (b) the cylindrical one for comparison. The orbital gaps are split and the gap between
corner and side states for the hexagonal geometry is reduced, yielding to shorter plateaus at
$12G_{\mathrm{0}}$.
Here $t= 0.2R_{\rm ext}=6$ nm}
\label{fig_ele}
\end{figure}

In these examples we assumed the geometric symmetry of the nanowires.
Nevertheless, shells with perfect polygonal symmetry cannot be
fabricated, such that the symmetric case should be considered only for
reference.  Geometric imperfections, random impurities, or external electric fields
(gates) remove the orbital degeneracies.  In order to account for asymmetry we consider 
an electric field transversal to the wire, included in Eq.\ \ref{hamiltonian_t}.
In practice, this is a way to model also the effect of a lateral gate attached to 
the nanowire, or of the substrate where the nanowire is situated, typically used to control
the carrier concentration with a voltage \cite{Sitek15}.  
Assuming this voltage is not very large, 
such that the electrons are still distributed over the entire shell (and not 
simply crowded near the gate), new plateaus can be obtained between
the split orbital states, as shown in Fig.\ \ref{fig_ele}, where we compare the results
obtained for a cylindrical and a hexagonal nanowire. 

The electric field changes the electron density distribution by favoring 
some corner or sides over the others, depending on its orientation. This
allows for the controllability of the electron localization within the
shell. Although the first  plateaus are similar for both the cylindrical and the hexagonal case the main difference arises from the orbital degeneracy pattern of the hexagonal nanowire, which is split at $12G_{\mathrm{0}}$, separating corner from side states (2442/2442). It is also interesting to observe
that, in the cylindrical nanowire, the electric field applied needs to be progressively
stronger in order to split the orbital degeneracy of higher energy levels, as shown
in Fig.\ \ref{fig_ele}(b).  Additionally, for the hexagonal wire, rotating the direction of the electric
field leads to a change in the secondary plateaus but leaves the main one
(at $12G_0$) unaffected (not shown).

\section{Effects of a longitudinal magnetic field}

We consider now a magnetic field in Eq.\ (\ref{hamiltonian_t}), 
longitudinal to the nanowire and its effects on the transverse modes.  
For a circular nanowire with $t \ll
R_{\rm ext}$ a periodic energy spectrum vs. the magnetic
field is expected, with period of one flux quantum in the cross sectional area.
Such flux-periodic oscillations, related to
the Aharonov-Bohm interference, have been observed experimentally on
InAs/GaAs hexagonal core-shell nanowires with $t=25$ nm and $R_{\rm
ext}\approx 75$ nm \cite{Gul14}.  The oscillations had additional
modulations that were attributed to impurities, or subsequently to the
spin splitting, by using a circular nanowire model of zero thickness 
\cite{Rosdahl14}.  

\begin{figure}[t]
 \centering
 \includegraphics[scale=0.67]{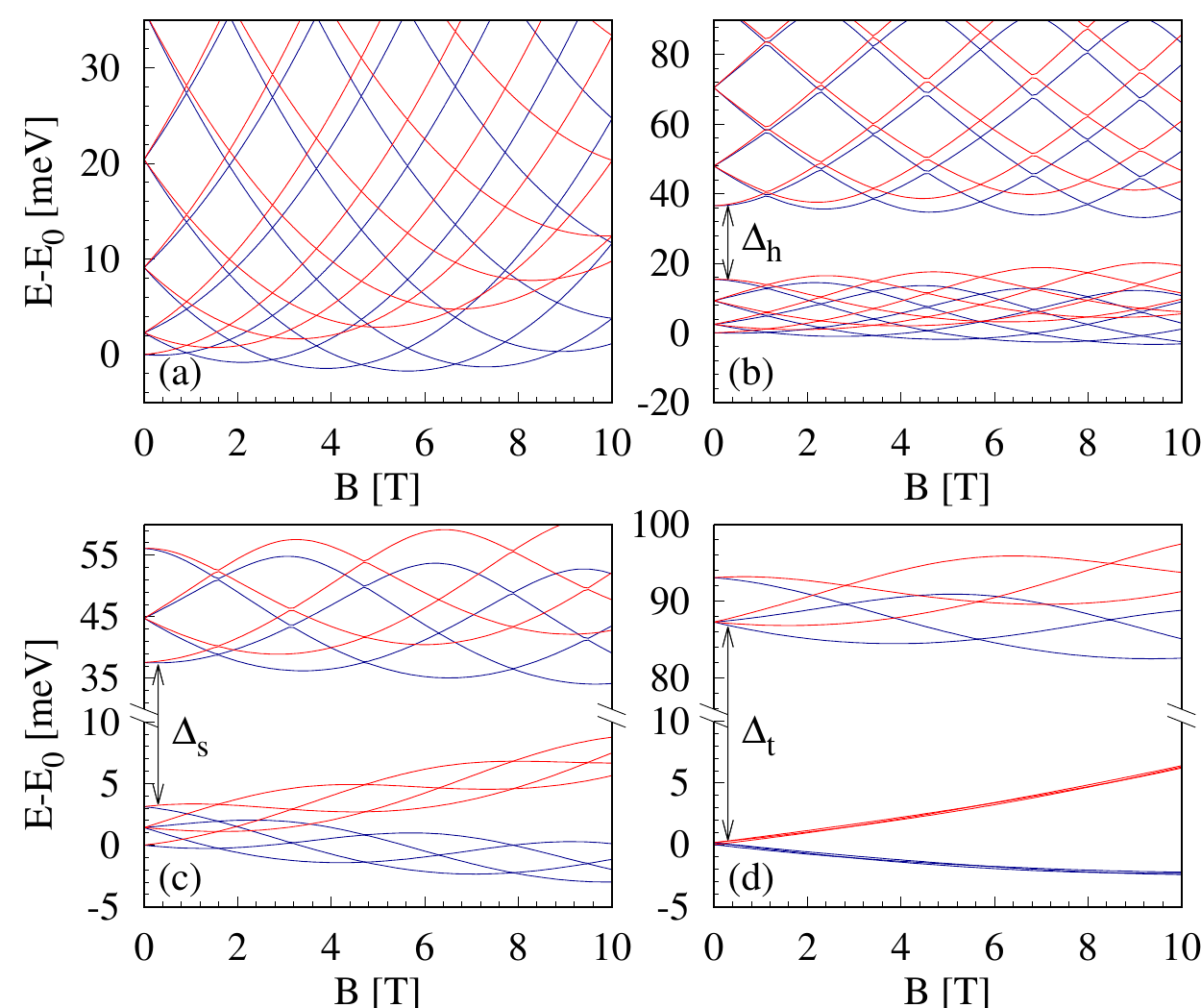}
 \caption{Evolution of the transverse energy modes with magnetic field 
 for the circular shape (a) and for the three polygonal shapes: (b) hexagonal, 
 (c) square and (d) triangular. The red color stands for the spin down levels 
 and the blue for the spin up ones.
Here $t=0.2R_{\rm ext}$.  Note that, in order to include both corner and 
side states in the graphs, the vertical axes in panels (c) and (d) are broken.}
\label{fig_transverse}
\end{figure}

Indeed, the presence of the spin disturbs or breaks the flux periodicity
of the transverse modes \red{when the magnetic field increases}, as seen
in Fig.\ \ref{fig_transverse}(a) for a circular shell.  \red{In addition, the
flux periodicity is also disturbed by the shell thickness. Flux periodic 
energy spectra for thin polygonal shells, without spin effects, 
have already been obtained by other authors \cite{Ferrari09a,Ferrari09b,Ballester13}}.

In Fig.\ \ref{fig_transverse}(b) we show the energies of the 
transverse modes for our hexagonal shell.  The energy gap between the corner and side
states decreases when the magnetic field is increased, but, remarkably, still a large value 
survives at high magnetic fields, in our case above 10 T.  The reason is that the orbital 
energy of the states localized in corners is significantly reduced compared
to states of higher energies, an effect that becomes more pronounced in the square 
geometry, Fig.\ \ref{fig_transverse}(c). 
\red{In fact, the differences reveal the symmetry reduction from circular
to hexagonal or square shapes.
In these examples the orbital degree of freedom of the corner states
is provided by their mutual coupling via tunneling across the polygon
sides. Instead, in the triangular case, where the corner localization is
the strongest, the tunneling is suppressed, and the orbital motion of the
corner states is completely frozen, such that only the spin Zeeman energy
is observable for the corner states in Fig.\ \ref{fig_transverse}(d). 
The same can also happen for thinner square and hexagonal shells.}

The conductance steps expected in the presence of a longitudinal magnetic
field would develop according to the evolution of the transverse modes
when increasing the field, by lifting both spin and orbital degeneracies.
It is worth noting that, up to 10 T, the gap separating corner and
side states prevails for the three polygonal shells, with the geometric
parameters used, and that a new gap separating the corner states with
different spin is created as a consequence of the Zeeman splitting for
the triangular and square nanowires, where the orbital degree of freedom
is most restricted.

\section{Effects of a magnetic field perpendicular to the nanowire}

\subsection{Charge and current distributions}

A magnetic field orthogonal to the nanowire axis, which
can be incorporated in the longitudinal Hamiltonian, Eq.\
(\ref{hamiltonian_l}), creates a second localization mechanism,
in addition to the localization imposed by the polygonal geometry.
This  type of magnetic localization has been studied in recent
years for the cylindrical shell geometry by several authors
\cite{Tserkovnyak06,Bellucci10,Manolescu13,Rosdahl15,Chang17}.  If the
electrons are confined on a cylindrical surface, their orbital motion
is governed by the radial component of the magnetic field.  Assuming a
magnetic field uniform in space, its radial component vanishes and
changes sign along the two parallel lines on the cylinder situated at
$\pm 90^{\circ}$ angles relatively to the direction of the field, and if
the field is strong enough the electrons have snaking trajectories along
these lines.  At the same time, at angles $0^{\circ}$ and $180^{\circ}$
the electrons perform closed cyclotron loops.  This localization mechanism
leads to an accumulation of electrons on the sides of the cylinder, where
the snaking states are formed, and to the depletion the regions hosting
the cyclotron orbits \cite{Manolescu13}.

\begin{figure}  [!h]
 \centering
\includegraphics[scale=0.4]{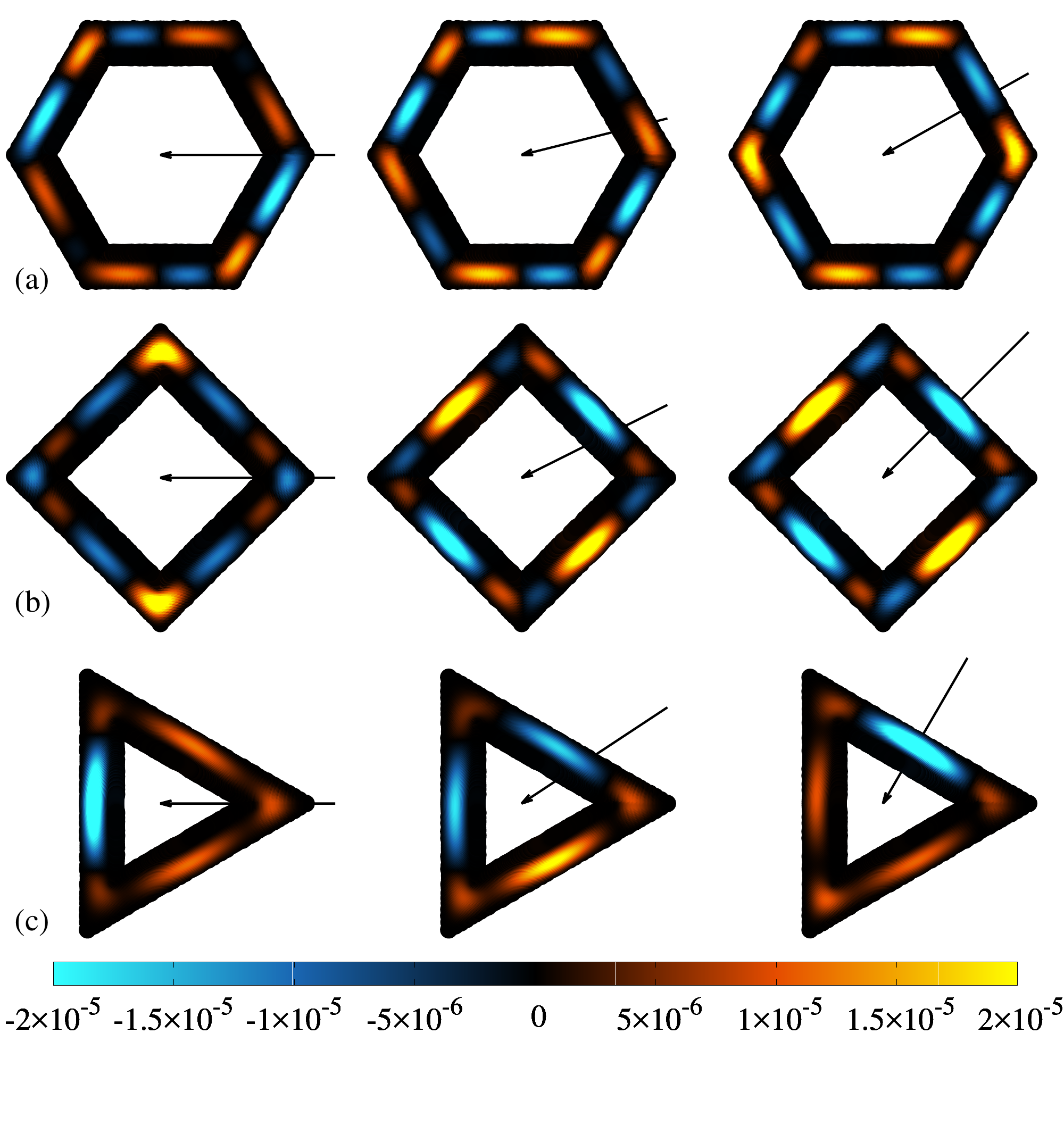}

\vspace{-5 mm}
\caption{The color scale shows the change of the electron density, in units
of nm$^{-3}$, to
the orientation of the magnetic field, of magnitude $B=2$ T, indicated
by the arrows, relatively to the three polygonal shells: (a) hexagonal,
(b) square, (c) triangular.  The magnetic field points to the corners in
the left column, it is perpendicular to the sides in the right column,
and half way in between these two orientations in the middle column.
In each case the electron density for $B=0$ is subtracted.  The average
electron density is fixed to $1.3 \times 10^{-4} \ {\rm nm}^{-3}$. 
The geometry parameters are $R_{\rm ext}=50$ nm and $t=10$ nm.
}
\label{2localiz}

\vspace{5 mm}
 \includegraphics[scale=0.4]{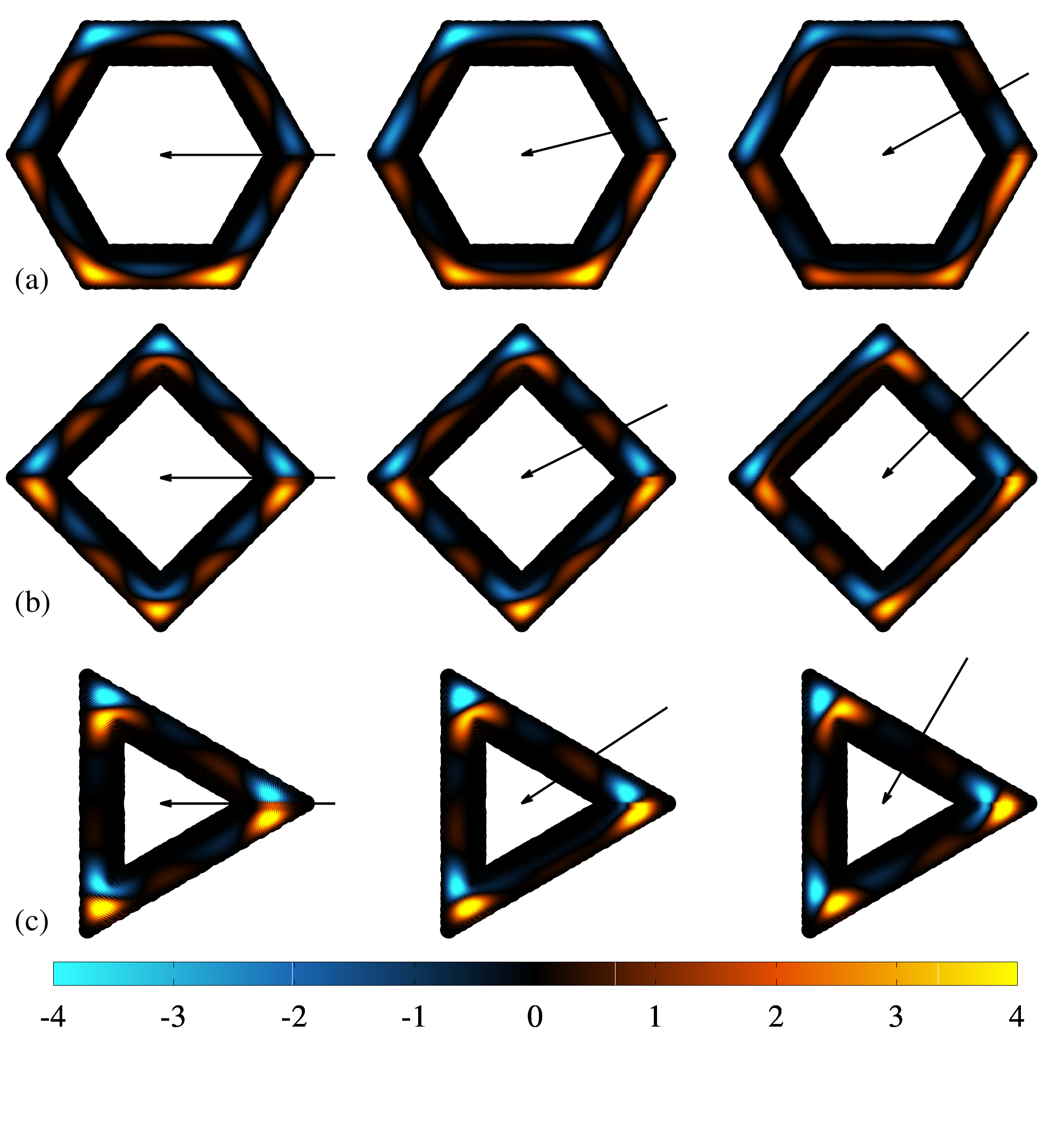}

\vspace{-5 mm}
\caption{The current density in equilibrium (no potential bias) for the same 
cases as in Fig.\ \ref{2localiz} above. The color scale is in units of nA/nm$^2$}
\label{2localizcurrent}
\end{figure}

In a thin prismatic shell the geometric and magnetic localization
of the electrons coexist.  Therefore, the distribution of electrons is
expected to depend on the orientation of the magnetic field relatively to the 
prism edges or facets, as in the examples shown in Fig.\ \ref{2localiz}. 
Here we consider a carrier concentration of $1.3 \times 10^{11} \ {\rm cm}^{-3}$ 
as reported in the recent literature \cite{Heedt16}.  In each case the color 
scale represents the difference between the carrier concentration at $B=2$ T 
and at $B=0$ for the three geometries.  With the present parameters 
these differences are somewhere up to 10\%.  Furthermore, for each geometry, 
the local changes of the density when the angle of the magnetic field
is varied, are up to 1\% only.  
Still, as we shall see, this variation should be sufficient to have implications
on the transport data.

Next, in Fig.\ \ref{2localizcurrent} we show the current distributions in the
equilibrium states, i.e. in the absence of a biased chemical potential ($\mu_+=\mu_-$).    
In the absence of the magnetic field the current distribution is zero everywhere
in the shell, i.e. the positive currents and negative currents compensate in
each point.  In the presence of the magnetic field, as expected, the Lorentz force
may create local currents, as shown in the figure, which no longer compensate
locally, but indeed the integrated current is zero.  The local currents are in fact 
loops along the $z$ axis of the nanowire, closing up at infinity. 
Still, it is interesting to observe the compensation of these loops. In
the cases where the magnetic field points along the direction of one
of the symmetry axes of the cross section the loops are compensated,
with the same current flowing in both directions, and the channels are
paired  with the ones on the opposite side of the geometric symmetry
axis relative to the magnetic field direction. The pairing can happen
within the same corner or side or on opposite ones. In the cases where the
magnetic field does not point along one of the symmetry axes the loops
are no longer compensated and the current traveling in both directions
is not the same. Instead, the compensation occurs when adding the loop
on the opposite side of the shell. 

\subsection{Energy spectra and conductivity}

\begin{figure}
 \centering
 \includegraphics[scale=0.43]{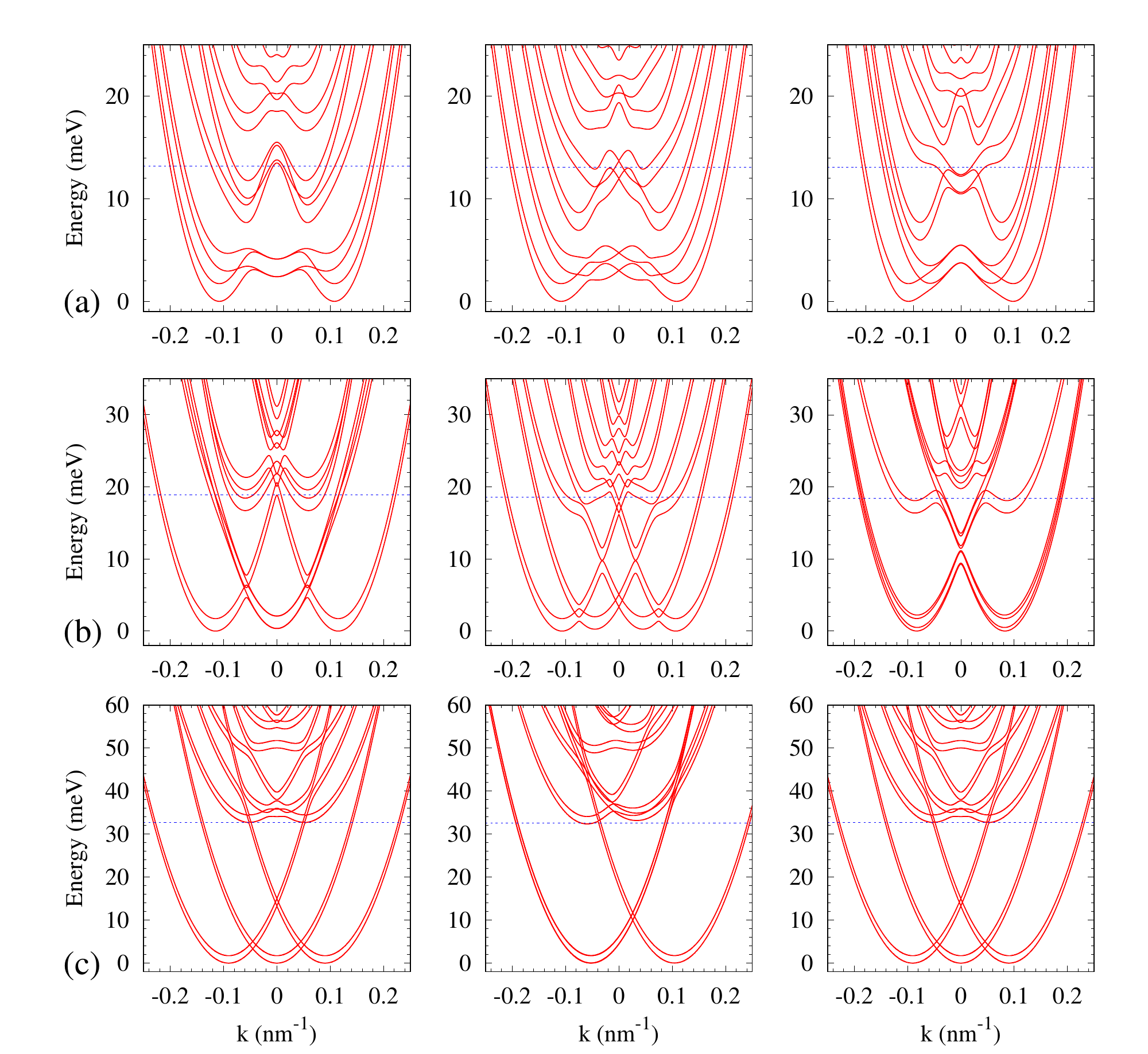}
\caption{Energy spectra obtained 
by varying the angle of the perpendicular magnetic field, as in Fig.\ \ref{2localiz}: 
(a) hexagon, (b) square,  (c) triangle.  In each case the energy of the 
ground state is considered zero. The temperature is 1 K.}
\label{speBperp}
\end{figure}

The energy dispersion with respect to the wave vector $k$, corresponding
to the situations shown in Figs. \ref{2localiz} and \ref{2localizcurrent},
can be seen in Fig.\ \ref{speBperp}.  The dashed horizontal lines
indicate the chemical potential corresponding to the selected carrier
concentration. In all cases, in the absence of the magnetic field
the position of the chemical potential is somewhere at the level of
the side states.  The presence of the magnetic field mixes the corner
and the side states and leads to quite complex changes in the spectra
and in the charge or current distributions.  And, as we can see, the
energy dispersions are also sensitive to the orientation of the field
relatively to the prismatic shell.  

Note that, in the triangular case, the spectra may not be symmetric
(even) functions of the wave vector $k$ if the magnetic field is not
aligned with a symmetry axis of the shell, like in the second example of
Fig.\ \ref{speBperp}(c), corresponding to the second example of Figs.\
\ref{2localiz}(c) and \ref{2localizcurrent}(c). In this case the magnetic field is parallel to the
side of the triangle.  The sign reversal of the magnetic field leads to
the sign reversal of the wave vector, i.e., to the same energy spectrum.
The reason for the asymmetric spectra is that the triangular geometry
is somehow special, compared to the hexagonal or square, because of the
absence of an inversion center.  

\begin{figure} 
 \centering
 \includegraphics[scale=0.64]{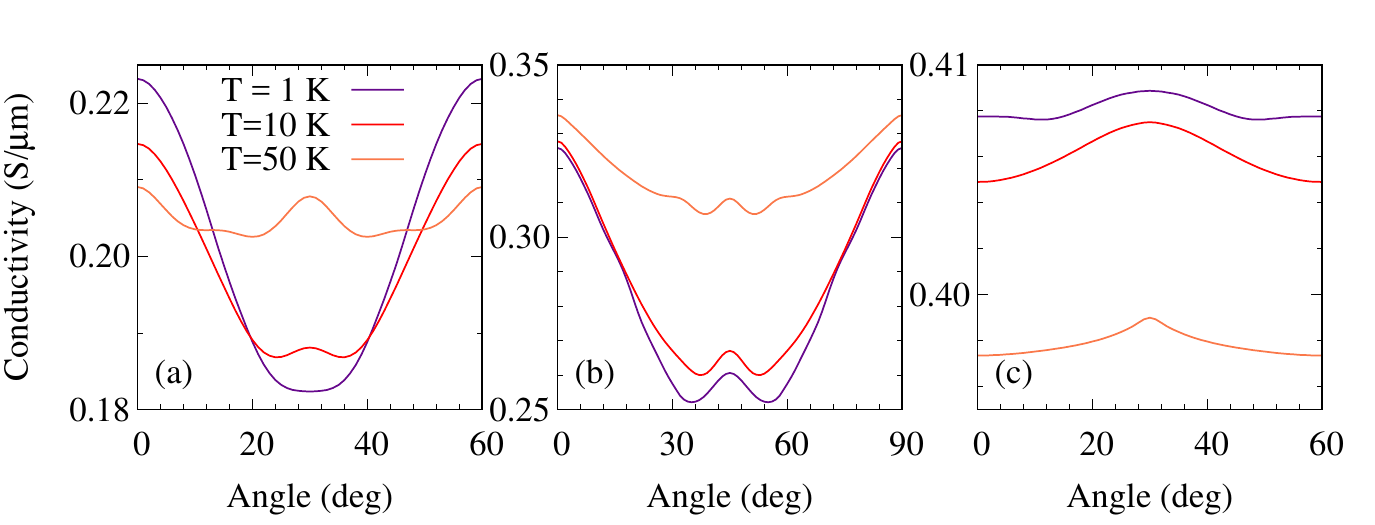}
\caption{Conductivity as a function of the angle of the magnetic field, 
for temperatures 1 K, 10 K, and 50 K, 
within one full period for each geometry
(a) hexagonal, (b) square, and 
(c) triangular shells. As before, the magnetic field is $B=2$ T 
and the carrier concentration $1.3 \times 10^{-4} \ {\rm nm}^{-3}$.
The results are obtained with Kubo formula including a disorder broadening
of 1 meV.
}
\label{aniscond}
\end{figure}

According to these results we can predict that due to the internal
geometry of the nanowire the conductance should depend on the orientation
of the magnetic field in a manner that indicates the polygonal cross section of
the shell.  We demonstrate this in Fig.\ \ref{aniscond} where
we show the results obtained with the Kubo formula (\ref{kuboc}). 
In these calculations we assume infinitely long nanowires and we include a 
disorder broadening of the energy spectra described by the parameter $\Gamma=1$ meV 
in Eq.\ (\ref{spectral_f}), i.e., we are far from the ballistic regime.  
In addition we consider temperatures up to 50 K, to emphasize that the conductance
anisotropy should be also robust to thermal perturbations. 

For each geometry the zero angle is considered when the field is
oriented towards a prism edge, as shown in the first column of Fig.\
\ref{2localiz}.  The angular period of the conductivity, when the
magnetic field rotates, is obtained when the magnetic field points
to the next corner for the hexagon and square, i.e.\ $60^{\circ}$ and
$90^{\circ}$, respectively, whereas for the triangle it corresponds to a
half of it, i.e. $60^{\circ}$.  Obviously, all figures can be continued
by periodicity, up to a complete rotation of the magnetic field.

The conductivity described by the Kubo formula is, in this case, 
an example of band conductivity, i.e., it is directly related to the velocity
of carriers along the transport direction, like in the ballistic case, and
decreases when the disorder parameter $\Gamma$ increases \cite{Manolescu97}. 
The dependence on the temperature is more complicated.  In this calculations we
completely ignored the dependence on the temperature of the collisional 
broadening parameter $\Gamma$, which is a separate problem.  In our model 
the main effect of increasing the temperature is the population/depopulation  
of the states above/below the chemical potential, respectively, and to a lesser 
extent the variation of the chemical potential itself.  Because the 
energy curves are nonlinear functions of the wave vector, Fig.\ \ref{speBperp}, 
the velocity distribution
of the carriers changes in a nonuniform manner, such that our conductivity may 
either increase or decrease with increasing the temperature.  
The relative variation of the conductivity with the angle is also not simple,
meaning that it can increase or decrease, depending on how the energy spectra behave around
the chemical potential when the magnetic field is changing orientation.  In our 
examples we can see that the variation is the weakest for the triangular case, which is
because the corner localization is stronger and their mixing with the side states
is relatively weak.

The conductance anisotropy in hexagonal core-shell nanowires was
already predicted for the case with electrons localized in the core
\cite{Royo13}, but only in the ballistic regime, at much higher magnetic
fields, and much less pronounced.  
\red{In our case the results obtained in the ballistic regime are qualitatively 
similar to those shown in Fig.\ \ref{aniscond}, except at low temperatures, when the
conductance depends critically on the number of intersections of the Fermi energy with
the energy bands, which may vary discontinuously. But, since it is
difficult to obtain the ballistic regime in realistic experimental
samples, our} results show that the anisotropy of the shell could be
easier resolved, in well attainable experimental conditions, and possibly
not only at low temperatures.

\subsection{Nonlinear I-V characteristics}

\begin{figure} [h]
 \centering
\includegraphics[scale=0.64]{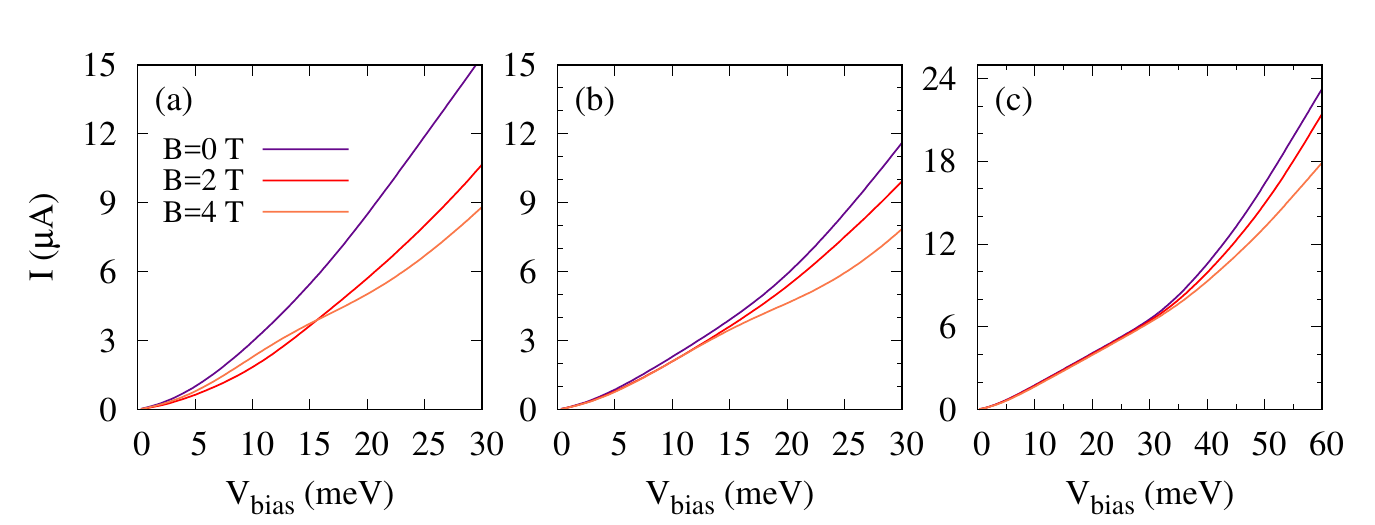}
\caption{$I-V$ characteristic for different values of a magnetic field perpendicular 
to an edge of the nanowire, for the three geometries: (a) the hexagonal, (b) square, and (c) triangular. Here
the temperature is 25 K.
The geometry parameters are $R_{\rm ext}=50$ nm and $t=10$ nm.}
\label{IVchar}
\end{figure}

Another indication of the internal geometry of the prismatic shell
can be found in the $I-V$ characteristic, as we show in Fig.\ \ref{IVchar}.
The $I-V$ characteristics are obtained following the same method used
for the conductance, i.e. Eq.\ (\ref{expected_current}).  This time we
create an imbalance between states with positive and negative velocity by
increasing the potential bias far from the linear regime, starting with
a chemical potential close to the lowest band bottom. 

We first notice the bias threshold where the linear regime breaks.  That is the
lowest for the hexagonal geometry, below 5 meV,
whereas for the square and triangular 
cases it increases to 15 meV and 35 meV, respectively. 
The change of slope of the curves, more clearly seen for square and
triangular cases,  indicate the transition from corner states to side
states, as more channels are crossed, and could offer a possibility to
observe experimentally the existence of the gap separating these groups
of states. In some cases the magnetic field yields a reduction in energy
and a shift in k-space of two mixed-spin bands, leading to an avoided
crossing at $k = 0$. In these cases an inflection point is observed,
more clearly for the hexagonal and square cases, as a consequence of
the reduction in the number of levels crossed by the potential bias.

In the results shown in Fig.\ \ref{IVchar} we used a temperature 
of 25 K in order to increase the population of the states above the chemical 
potential.
The $I-V$ curves were obtained assuming an infinite wire without impurities.  
To a first approximation their averaged effect can be incorporated in 
Eq.\ (\ref{expected_current}) by the substitution
\begin{equation}
\nonumber
{\cal F} \left(\frac{E_{mks}-\mu}{k_{B}T}\right) \rightarrow
\int A_{mks}(E) {\cal F} \left(\frac{E-\mu}{k_{B}T}\right)  dE \ .
\end{equation}
In this way we could obtain results similar to those shown in Fig.\ \ref{IVchar}
with a lower temperature, $T=1$ K, but with a disorder energy $\Gamma=1$ meV.


\section{\label{sec_conclusions} Conclusions}

We presented several features of the conductance of core-shell
nanowires with a conductive shell and an insulating core, which are
consequences of the internal geometry of such nanowires, and can be probed
in well achievable experimental conditions.  The motivation
of our work is to stimulate the interest of the experimental groups to
do such investigations and to achieve the corresponding quality of the
samples with a clear manifestation of the internal geometry.  

Most of the presented results are basically determined by the
energy spectra and by the geometric localization of the electrons.  In the ballistic
cases, with no impurities, there is no real need for transport calculations, 
the conductance can be obtained by simply counting the transverse modes.
The transport calculations that we performed were intended to support the 
predictions from the spectra, in a qualitative manner. We used the 
transmission function for short, ballistic or quasi-ballistic wires, 
and Kubo formula for long (infinite) non-ballistic wires.

One of the most interesting aspects is that the states localized along  
the edges and those localized on the facets can be separated by a
large energy gap, robust to many kinds of perturbations, such that a
single core-shell nanowire may possibly function like a collection of
thinner nanowires.  Another interesting result is that the localization,
and thus the conductance features, can be modified or controlled by
external electric or magnetic fields.

\begin{acknowledgments}
Instructive discussions with Patrick Zellekens, Thomas Sch\"apers, and Mihai Lepsa 
are highly appreciated.  This work was supported by the Icelandic Research Fund.
\end{acknowledgments}


\bibliographystyle{apsrev4-1}
\bibliography{core_shell}

\end{document}